\documentclass[aps,pra,superscriptaddress,twocolumn]{revtex4-1}
\usepackage[utf8]{inputenc}
\usepackage[english]{babel}
\usepackage{amsmath}
\usepackage{amsfonts}
\usepackage{amssymb}
\usepackage{graphicx}
\usepackage{amsthm}
\usepackage{color}
\usepackage{hyperref}
\usepackage{paralist}
\newcommand{\ket}[1]{\ensuremath{\left|#1\right\rangle}}
\newcommand{\bra}[1]{\ensuremath{\left\langle #1\right|}}

\newcommand{\dbraket}[2]{\ensuremath{\langle #1 |#2 \rangle}}
\newcommand{\ketbra}[1]{\ensuremath{| #1 \rangle\langle #1 |}}
\newcommand{\dketbra}[2]{\ensuremath{| #1 \rangle\langle #2 |}}
\newcommand{\abs}[1]{\ensuremath{\left|#1\right|}}

\begin{document}
\title{Multi-Phase Hadamard receivers for classical communication\\ on lossy bosonic channels} 
\author{Matteo Rosati}
\affiliation{NEST, Scuola Normale Superiore and Istituto Nanoscienze-CNR, I-56127 Pisa,
Italy.}
\author{Andrea Mari} 
\affiliation{NEST, Scuola Normale Superiore and Istituto Nanoscienze-CNR, I-56127 Pisa,
Italy.}
\author{Vittorio Giovannetti} 
\affiliation{NEST, Scuola Normale Superiore and Istituto Nanoscienze-CNR, I-56127 Pisa,
Italy.}

\begin{abstract}
 A scheme for transferring classical information over a lossy bosonic channel is proposed by generalizing the proposal presented in  Phys. Rev. Lett.  \textbf{106}, 240502 (2011) by Guha. It employs
 codewords  formed by products of coherent states of fixed mean photon number with multiple phases which, 
through a passive unitary transformation, reduce  to  a Pulse-Position Modulation code with multiple pulse phases. 
The maximum information rate achievable with optimal,  yet difficult to implement, detection schemes is computed and 
   shown to saturate the classical capacity of the channel in the  low energy regime. An easy to implement receiver  based on a conditional  Dolinar detection scheme is also proposed finding that, while suboptimal, it allows for  improvements  in an intermediate photon-number regime with respect to previous proposals.
\end{abstract}
\maketitle

\section{Introduction}
The field of optical communications is one where Quantum Information may have its first promising applications, thanks to its somewhat easier technological requirements than, for example, computation. Several theoretical works in the past have produced the ultimate bound on the transmission of classical information through a quantum channel, i.e., the classical-quantum capacity, and demonstrated its achievability \cite{HolevoBOOK,holevo1,holevo2,schumawest,holevo3,winter,oga,hauswoot,oganaga,hayanaga,hayashi,seq1,seq2,sen,hastings}; in particular, recently it has been shown that the classical-quantum capacity for gaussian channels, which are used to model the most common communication media, is achieved by a gaussian encoding \cite{gaussOpt,maj1,maj2}. Nevertheless there is still much work going on to achieve that capacity in practice \cite{wildeguha1,wildeguha2,takeokaGuha,takeokaGuha2,lee}. This is mainly due to the fact that optimal decoding strategies involve joint measurements on long codewords of quantum states \cite{schumawest,hauswoot,seq1,seq2,sen,polarWildeGuha,Arikan,wildeHayden,NOSTRO}, which are difficult to implement with gaussian operations and photodetectors. 

A recent proposal by Guha~\cite{Guha1}, which we call the Hadamard receiver, points in this direction, representing one of the first structured joint-detection receivers for optical communications, realizable in principle with current technology \cite{Banaszek}. This detection scheme relies on building a codebook  (Hadamard code in the following) of $n$ separable codewords of length $n$ from the binary coherent alphabet $\ket{\pm\alpha}$ and transforming them at the receiver by a passive unitary  gate $\hat{U}_{Had}^{(n)}$  represented by a Hadamard matrix \cite{Hadamard}. The output is a Pulse-Position-Modulation (PPM) code, which is easily read out by employing single-mode quantum state discrimination techniques, e.g., photodetection. In the low-energy regime the scheme surpasses the   information rate obtained by separable detection techniques, jointly reading out larger and larger codewords as the energy gets lower. Further improvements can also be gained~\cite{Guha2} 
by adding a second copy of the original Hadamard  codebook obtained from the latter  by simply flipping the sign of the amplitudes  of all the coherent states that form its codewords:
the new vectors behave 
properly under the Hadamard transformation $\hat{U}_{Had}^{(n)}$, producing  two (phase-shifted) copies of the initial PPM code  which can be still read out  by means of an adaptive Dolinar receiver~\cite{Dol,holDol,ImplProj,Multicopy,DolMultiplexed,DolExp,DolImp}.
Building up from these observations here 
 we analyse in detail the case of a codebook formed by 
 $M$  phase-shifted copies of the Hadamard code (the case $M=2$ corresponding to the model discussed in Ref.~\cite{Guha2}): we call this a 
 Phase-Shift-Keying (PSK) Hadamard  code of order $M$. 
First, by  exploiting the symmetries of the problem, we compute  the optimal  information rate  of this code, evaluating its associated Holevo information. We show that, for all choices of the number  of phase modulations $M\geq 2$ and of the codewords' length $n$,  the optimal rate saturates the classical capacity of the channel~\cite{HolevoBOOK} in the low-energy region. Accordingly in such regime the PSK Hadamard codes are optimal  and any sub-performance resulting from their use is only a consequence of a lack of efficiency at the detection stage.   Next we  compute  the rates attainable when detecting PSK Hadamard codes with   a modification of the Hadamard receiver of Ref.~\cite{Guha2}, which we dub  PSK Hadamard receiver. It 
  relies on  the discrimination  of multiple symmetric coherent states of fixed intensity~\cite{multiHel1,multiHel2,bondurant,izumi,guhaDol,becerra1,becerra2,marquardt,Hel}, applied to the PPM vectors  
 that emerge from the Hadamard transformation $\hat{U}_{Had}^{(n)}$, operating on the transmitted elements of the PSK Hadamard code. 
  The rate attainable by means of this technique  saturates the optimal rate of the code at high energy, but it levels off at a plateau at low energy. Comparing the results obtained
  for different  values  of $M$ we observe also that, while in the regime of  low photon numbers  $M=2$ appears to be the best choice, the use of more than two phases yields better performances in a crossover energy region where separable techniques start to perform worse than the one described in Refs.~\cite{Guha1,Guha2} (the relative  improvement being up to $6\%$ with respect to the $M=2$ case). 

The manuscript is organized as follows: in Sec. \ref{sec2} we introduce the formal definition of the PSK Hadamard code and its properties. Then in Sec. \ref{secHadCap} we compute its associated optimal rate.  
The performances attainable by means of the PSK Hadamard  receiver are presented instead in Sec. \ref{secHadRate}. The paper ends with Sec.~\ref{secConc}, where we discuss our results and draw some conclusions.

\section{PSK Hadamard codes}\label{sec2}
Here we formalize the  Hadamard  code proposed in Ref.~\cite{Guha1} and its $M$-phase generalization. 
Suppose we want to transfer classical information through $n$ modes of the electromagnetic field, or $n$ distinct temporal pulses on a single field mode, traveling through a communication medium, e.g., an optical fiber or free space~\cite{CAVES}. The transmission line can be approximately represented by a phase-insensitive lossy bosonic channel of loss $\eta$, which transforms the single-mode bosonic annihilation operator $\hat{a}$ of the field as $\hat{a}^{'}=\sqrt{\eta}~\hat{a}+\sqrt{1-\eta}~\hat{e}$,  with $\hat{e}$ the annihilation operator
of the channel environment which is assumed to be in its vacuum state.
It is well known~\cite{EXACT} that, by imposing an upper bound  $E$ on the average photon number  that we are allowed to use per mode, the classical information  capacity of this channel~\cite{HolevoBOOK} can be achieved, in the limit of large $n$,  by means of  codewords formed by products of coherent
inputs whose amplitudes are sampled from a Gaussian distribution centered in zero and with variance $E$.  Ultimately this is possible because when we encode information in the optical coherent states of the field, $\ket{\alpha}$, 
they reach the receiver end of the channel
as attenuated versions $\ket{\sqrt{\eta}~\alpha}$, while keeping their original purity.
Since the net action of the channel on these states is a rescaling of energy by $\eta$, we may take as a reference the \textit{received} energy without loss of generality, which is equivalent to setting $\eta=1$ in the following.

 The practical use of the Gaussian code described above is 
limited by the difficulty in implementing  the optimal detection scheme that is able to read its codewords efficiently. This motivates the search for alternative ways of encoding classical  messages into the channel which, while being possibly not as efficient as the Gaussian one, will guarantee nevertheless better performances with readout strategies which are
easier to implement.  The proposal of Ref.~\cite{Guha1} is one of such attempts. As in the Gaussian case  it employs sequences of coherent states as codewords. At variance with the latter however
such sequences are selected by extracting elements from a binary coherent alphabet  $\ket{\pm\alpha}$,  with amplitude  $\alpha$
matching the average-energy constraint of the channel, i.e. $|\alpha|^2 = E$.   
The number of different $n$-long strings one can generate this way is equal to $2^n$ (specifically they are the vectors 
$\ket{\pm \alpha}_{0}\otimes\cdots\otimes\ket{\pm\alpha}_{n-1}$ where $\ket{\cdot}_{j}$ represents the state of the $j$-th communication mode). 
  Yet in constructing the Hadamard code we only select $n$ specific ones of them
 by picking those sequences  whose signs reproduce the columns of a Hadamard matrix \cite{Hadamard}.  For instance for $n=2$  the Hadamard matrix, up to a normalization, is equal to  $H_{2}=\left[\begin{array}{cc}+1&+1\\+1&-1\end{array}\right]$: therefore 
  from the set of four possible elements 
  we select $\ket{v_{0}(\alpha)}=\ket{+\alpha}_{0}\otimes\ket{+\alpha}_{1}$ and $\ket{v_{1}(\alpha)}=\ket{+\alpha}_{0}\otimes\ket{-\alpha}_{1}$ as
  first and second codeword of our Hadamard codebook. 
For arbitrary $n$ we recall that  a Hadamard matrix of order $n=2^{i}$  with integer $i\geq0$, is a $n\times n$ matrix $H_{n}$ of elements $\pm 1$ which is orthogonal up to a scaling factor of $\sqrt{n}$ and, for simplicity, symmetric (a permutation of columns or rows is sufficient to satisfy this further requirement), i.e., $H_{n}H_{n}^{T}=H_{n}^{T}H_{n}=n\mathbf{1}_{n}$, $H_{n}^{T}=H_{n}$~\cite{Hadamard}. Such a matrix can be equivalently defined in terms of its elements as:
\begin{equation} \label{HadTerm}
(H_{n})_{j,k}=(-1)^{j\cdot k},\quad j\cdot k=\sum_{t=0}^{\log_{2}n}j_{t}k_{t},
\end{equation}
where $j\cdot k$ is the bitwise scalar product of the binary representations of $j,k=0,\cdots,n-1$.
Then a Hadamard code ${\cal H}_1^{(n)}(\alpha)$ of length $n$ and average energy per mode $E=\abs{\alpha}^{2}$ comprises the  $n$ codewords
\begin{eqnarray} \ket{v_{k}(\alpha)}=\bigotimes_{j=0}^{n-1}\ket{(H_{n})_{j,k}~\alpha}_{j}\;,\end{eqnarray} for all $k=0,\cdots,n-1$.
As shown in Refs. \cite{Guha1,Guha2,Banaszek}, this special set of states admits a   passive unitary transformation $\hat{U}^{(n)}_{had}$, completely implementable through passive optical elements as beam splitters and phase shifters, that transforms the received codewords $\ket{v_{k}(\alpha)}$ into equivalent PPM ones, i.e., 
\begin{eqnarray}
\hat{U}^{(n)}_{had} \ket{v_{k}(\alpha)} = \ket{w_{k}(\alpha)}=\ket{\sqrt{n}~\alpha}_{k}\otimes\left(\bigotimes_{j\neq k}\ket{0}_{j}\right), \label{TRANS1}
\end{eqnarray} characterized by a single high-energy pulse on one of the $n$ available modes, the vector $|0\rangle_j$ representing the vacuum state of the $j$-th mode 
 (see Appendix~\ref{APPA} for more details on $\hat{U}^{(n)}_{had}$).
Note that the average received energy per mode $E$ is conserved by the transformation, although its total amount for $n$ modes is concentrated on a single pulse of energy $\mathcal{E}=nE$, whose position varies from codeword to codeword. Accordingly the classical messages encoded into the elements of ${\cal H}_1^{(n)}(\alpha)$ can now be recovered by means of a readout strategy capable of detecting where such concentrated energy lies, e.g. photodetection or a Dolinar receiver~\cite{Dol,holDol,ImplProj,Multicopy,DolMultiplexed,DolExp,DolImp} aimed to discriminate the coherent state $|\sqrt{n} \alpha\rangle$ from the vacuum  $|0\rangle$.

As already mentioned in the introduction, a refined version of the receiver proposed in Ref.~\cite{Guha2} employs an enlarged code, obtained by adding those codewords $\ket{v_{k}(-\alpha)}$ with signs opposite to the ones in $\ket{v_{k}(\alpha)}$. More generally, we define a PSK Hadamard code of order $M$ 
by adding $M$ copies of  the Hadamard code, characterized by $M$ equally spaced phases  of the amplitude $\alpha$, i.e. the set 
\begin{eqnarray} \label{PSKHAD} 
{\cal H}^{(n)}_M(\alpha) &=& \cup_{m=0}^{M-1}\; {\cal H}^{(n)}_1(\alpha_{m})\;,
\end{eqnarray} 
 where $\alpha_{m}=e^{i\frac{2\pi}{M}m}\alpha$. Such code is 
 the collection of $Mn$ vectors
$\ket{v_{k}(\alpha_{m})}$ 
with  $k=0,\cdots,n-1$ and  $m=0,\cdots,M-1$. 
Since the property \eqref{TRANS1} holds for any $\alpha\in\mathbb{C}$, the elements of ${\cal H}^{(n)}_M(\alpha)$ get
 transformed by $\hat{U}^{(n)}_{had}$ in corresponding phase-shifted PPM codewords, i.e. the vectors
 \begin{eqnarray}
\hat{U}^{(n)}_{had} \ket{v_{k}(\alpha_m)} = \ket{w_{k}(\alpha_m)}\;. \label{TRANS2}
\end{eqnarray}
As we shall discuss explicitly in Sec.~\ref{secHadRate}, the readout of the input messages can benefit from this effect, the idea being to first
determine the value of $k$ by checking in which of the $n$ modes the energy has been concentrated, and then determine $m$ by using a single-mode detection scheme to read out the phase of $\alpha_m$.

\section{Optimal communication rates of the  PSK Hadamard codes}\label{secHadCap}
In this section we compute the optimal communication rate   $R^{(n,M)}_{opt}(E)$ of a PSK Hadamard code ${\cal H}^{(n)}_M(\alpha)$ of order $M$ and average energy per mode  $E=|\alpha|^2$. 
This  is proportional to the Holevo $\chi$-information \cite{HolevoBOOK} of the code itself. More precisely 
defining 
\begin{eqnarray}
\bar{\rho}=\sum_{k=0}^{n-1}\sum_{m=0}^{M-1}\frac{\ketbra{v_{k}(\alpha_{m})}}{M n}\;,\end{eqnarray}  the average state of the  code, we are interested into the quantity 
  \begin{eqnarray} 
R^{(n,M)}_{opt}(E) &=&\frac{1}{n} \left[ S(\bar{\rho})-\sum_{k=0,m=0}^{n-1,M-1}\frac{S(\ketbra{v_{k}(\alpha_{m})})}{M n}\right] \nonumber \\
 &=& S(\bar{\rho})/n\;, \label{CHIRATE}
\end{eqnarray} 
where $S(\cdot) = \mbox[ (\cdot) \log_2 (\cdot)]$ is the Von Neumann entropy of a quantum state \cite{NChuang}, and where the division by $n$ takes into account that we are interested
in the amount of information which can be transferred  \textit{per mode}.
 The quantity $R^{(n,M)}_{opt}(E)$ is the maximum rate of bits one can convey over the channel with the code  ${\cal H}^{(n)}_M(\alpha)$ with an optimal decoding procedure, e.g.~\cite{HolevoBOOK,seq1,seq2,schumawest,polarWildeGuha,NOSTRO}. 
To compute it we find it useful to exploit the unitary mapping~(\ref{TRANS2}). Accordingly the eigenvalues, and hence the entropy, of $\bar{\rho}$ coincide with those
of the density matrix 
\begin{eqnarray}\label{ENTROPY} 
\bar{\rho}_{PPM}&=& \sum_{k=0}^{n-1} \sum_{m=0}^{M-1}\frac{\ketbra{w_{k}(\alpha_{m})}}{M n} \nonumber \\
&=&
\frac{1}{n} \sum_{k=0}^{n-1}\bar{\rho}^{loc}_{k}\otimes\left(\bigotimes_{j\neq k}\ket{0}{}_{j}\bra{0}\right)\;, 
\end{eqnarray} 
with
\begin{eqnarray} 
\bar{\rho}^{loc}_{k}=\frac{1}{M} \sum_{m=0}^{M-1}\ket{\sqrt{n}\alpha_{m}}{}_{k}\bra{\sqrt{n}\alpha_{m}} \;.
\end{eqnarray} 

By construction the state $\bar{\rho}_{PPM}$  has non-zero support only on a $M n$-dimensional subspace of the whole Hilbert space, spanned by the linearly independent states  $\ket{w_{k}(\alpha_{m})}$: we compute the entropy of $\bar{\rho}_{PPM}$ by finding its spectrum in this subspace.
We proceed in two steps: first, for all $k$, we diagonalize $\bar{\rho}_{k}^{loc}$, then we diagonalize the resulting multimode superposition, which turns out to be already partially diagonal. The first step has already been carried out in Ref.~\cite{multiHel2}, when computing the Helstrom probability of discriminating the states $\{ |\alpha_m\rangle, m=0, \cdots, M-1\}$ (see also Sec.~\ref{secHadRate}). Accordingly we can write
\begin{eqnarray}  \label{QUESTA}
\bar{\rho}_k^{loc}=\sum_{\ell=0}^{M-1}\frac{\lambda_{\ell}(\mathcal{E})}{M}\ket{d_{\ell}}{}_{k}\bra{d_{\ell}} \;, \end{eqnarray} 
where the eigenvectors
\begin{equation}\label{mpskEigenvectors}
\ket{d_{\ell}}_k=\sum_{m=0}^{M-1}\frac{e^{-i\frac{2\pi}{M}\ell m}}{\sqrt{M\lambda_{\ell}(\mathcal{E})}}\ket{\sqrt{n}\alpha_{m}}_k\;,
\end{equation}
are obtained by Fourier transform of the original pulses $\ket{\sqrt{n} \alpha_{m}}_k$ and the eigenvalues
\begin{equation}\label{mpskEigen}
\lambda_{\ell}(\mathcal{E})=\sum_{h=0}^{M-1}\exp\left[-\left(1-e^{i\frac{2\pi}{M}h}\right)\mathcal{E}-i\frac{2\pi}{M}\ell h\right]
\end{equation}
depend on the fixed energy of the states, $\mathcal{E}=n E$ in our case.
It is important to note that the eigenvectors $\ket{d_{\ell}}_k$ for all allowed $\ell>0$ share the peculiar property of having a zero overlap with the vacuum state, i.e., 
\begin{eqnarray}\label{OVER}
{}_k\dbraket{0}{d_{\ell}}_k=\delta_{\ell, 0}~e^{-\mathcal{E}/2}\sqrt{M/\lambda_{0}(\mathcal{E})}\;. \end{eqnarray} This is due to the fact that the overlap of any PSK state with the vacuum depends only on its fixed energy and not on its specific phase, i.e., $\dbraket{0}{\sqrt{n}\alpha_{m}}=e^{-\mathcal{E}/2}$. Hence this overlap factors out of the sum in \eqref{mpskEigenvectors}, which then amounts to a simple discrete Fourier transform of a constant.
Replacing Eq.~(\ref{QUESTA}) into (\ref{ENTROPY}) we can hence write 
\begin{eqnarray}
\bar{\rho}_{PPM} = \sum_{\ell=0}^{M-1} \nu^{\ell}(\mathcal{E})  \sum_{k=0}^{n-1} \ketbra{e_{k}^{\ell}} \;, 
\end{eqnarray} 
with $\nu^{\ell}(\mathcal{E})=\lambda_{\ell}(\mathcal{E})/(M n)$ and 
\begin{eqnarray}
\ket{e_{k}^{\ell}}:=\ket{d_{\ell}}_{k}\otimes\left(\bigotimes_{j\neq k}\ket{0}_{j}\right)\;.\label{NEw1}\end{eqnarray} 
These new multimode states share the PPM structure of the $\ket{w_{k}(\alpha_{m})}$ but with different single-mode pulses, one for each eigenvector of $\bar{\rho}^{loc}_k$. Most importantly, thanks to the property~(\ref{OVER}) they turn out to be partially orthogonal, i.e. 
\begin{equation}\dbraket{e_{k}^{\ell}}{e_{h}^{m}}=\begin{cases}
\dbraket{d_{\ell}}{d_{m}}_{k}=\delta_{\ell, m} &\text{if }k=h,\\
\dbraket{d_{\ell}}{0}_{k}\cdot\dbraket{0}{d_{m}}_{h}=\delta_{\ell, 0}\delta_{m, 0} \frac{M e^{-\mathcal{E}}}{\lambda_{0}(\mathcal{E})}&\text{if }k\neq h.
\end{cases}\label{eOverlaps}
\end{equation}
It is then advantageous to write   $\bar{\rho}_{PMM}$ as the sum of two operators, $\bar{\rho}_{PPM}=\bar{\rho}^{(0)}_{PPM}+\bar{\rho}^{(+)}_{PPM}$, with
\begin{align}\label{rho0}
\bar{\rho}_{PPM}^{(0)}&=\nu^{0}(\mathcal{E})
\sum_{k=0}^{n-1}\ketbra{e_{k}^{0}},\\ \label{rho+}
\bar{\rho}_{PPM}^{(+)}&=\sum_{\ell=1}^{M-1}\nu^{\ell}(\mathcal{E})\sum_{k=0}^{n-1}\ketbra{e_{k}^{\ell}}\;.
\end{align}
From Eqs. (\ref{eOverlaps}-\ref{rho+}) we can deduce several facts: (i) $\bar{\rho}_{PPM}^{(0)}\bar{\rho}_{PPM}^{(+)}=\bar{\rho}_{PPM}^{(+)}\bar{\rho}_{PPM}^{(0)}=0$, i.e., the two operators are disjoint; (ii) the set of states $\ket{e_{k}^{\ell>0}}$ is an orthonormal basis of the space spanned by $\bar{\rho}^{(+)}_{PPM}$, of dimension  $(M-1)n$; (iii) $\bar{\rho}_{PPM}^{(+)}$ is diagonal in this basis, with eigenvalues $\nu^{\ell>0}(\mathcal{E})$ of multiplicity $n$ each. 
Accordingly to determine the spectral structure of $\bar{\rho}_{PPM}$ we have to diagonalize $\bar{\rho}^{(0)}_{PPM}$. This can be carried out by diagonalizing the Gram matrix \cite{multiHel2} of the codewords $\ket{e_{k}^{0}}$, whose average state is proportional to $\bar{\rho}^{(0)}_{PPM}$ itself. Indeed define the Gram matrix of a set of codewords as the $n\times n$ hermitian matrix whose entries are the overlaps between the desired codewords, i.e., $(\Gamma^{(e)})_{k,h}=\dbraket{e_{k}^{0}}{e_{h}^{0}}$ in our case. Since such matrix is diagonalizable, there exist a diagonal matrix $D=\operatorname{diag}(\mu_{0},\cdots,\mu_{n-1})$ and a unitary matrix $V$ that satisfy $D=V\Gamma^{(e)}V^{\dagger}$, with $\mu_{j}$ the eigenvalues of $\Gamma^{(e)}$. Then we can construct a new set of codewords $\ket{f_{j}}=\frac{1}{\sqrt{\mu_{j}}}\sum_{k=0}^{n-1}(V_{j,k})^{*}\ket{e_{k}^{0}}$ with two peculiar properties: (i) they form an orthonormal basis of the space spanned by the $\ket{e_{k}^{0}}$, since $\dbraket{f_{j}}{f_{i}}=\frac{1}{\sqrt{\mu_{j}\mu_{i}}}\sum_{k,h=0}^{n-1}V_{j,k}(\Gamma^{(e)})_{k,h}(V^{\dagger})_{h,i}=\delta_{j,i}$; (ii) $\bar{\rho}^{(0)}_{PPM}$ is diagonal in this basis. The latter property can be easily shown by inverting the definition of the basis states, i.e., $\ket{e_{k}^{0}}=\sum_{j=0}^{n-1}V_{j,k}\sqrt{\mu_{j}}\ket{f_{j}}$, and inserting it in \eqref{rho0} to obtain 
\begin{eqnarray} \bar{\rho}^{(0)}_{PPM}&=&
\nu^{0}(\mathcal{E})
\sum_{k,j,i=0}^{n-1}\sqrt{\mu_{j}\mu_{i}}V_{j,k}(V_{i,k})^{*}\dketbra{f_{j}}{f_{i}}\nonumber \\
&=&
\nu^{0}(\mathcal{E})
\sum_{j=0}^{n-1}\mu_{j}\ketbra{f_{j}}.\end{eqnarray}  In Appendix~\ref{AppA} we report a detailed computation of the spectral decomposition of $\Gamma^{(e)}$. For our purpose here, it is sufficient to say that the Gram matrix is diagonalized by a unitary $V=H_{n}/\sqrt{n}$, i.e., by the Hadamard matrix of order $n$, defined in Sec.~\ref{sec2}; moreover the eigenvalues of $\Gamma^{(e)}$ are $\mu_{j}(\mathcal{E})=1+(n\delta_{j,0}-1)\frac{M e^{-\mathcal{E}}}{\lambda_{0}(\mathcal{E})}$. Thus the average state \eqref{rho0} can be written in diagonal form as
\begin{align}\label{rho0Fin}
\bar{\rho}^{(0)}_{PPM}=\nu_{0}^{0}(\mathcal{E})\ketbra{f_{0}}+\nu_{+}^{0}(\mathcal{E})\sum_{k=1}^{n-1}\ketbra{f_{k}},
\end{align}
with eigenvalues 
\begin{eqnarray} \nu_{0}^{0}(\mathcal{E})=\frac{1}{M n}[\lambda_{0}(\mathcal{E})+(n-1)M e^{-\mathcal{E}}]\;,
\end{eqnarray}  of multiplicity one and 
\begin{eqnarray} \nu_{+}^{0}(\mathcal{E})=
\frac{1}{M n}[\lambda_{0}(\mathcal{E})-M e^{-\mathcal{E}}] \;, \end{eqnarray} of multiplicity $n-1$.
We are now in the position of computing the optimal rate~(\ref{CHIRATE}). This is 
\begin{eqnarray}\label{mpskHolCap}
&&R^{(n,M)}_{opt}(E)=S(\bar{\rho}_{PPM})/n= 
-\frac{1}{n}\Bigg[\nu_{0}^{0}\left(\mathcal{E}\right)\log_{2}\nu_{0}^{0}\left(\mathcal{E}\right)\\
&&\nonumber +(n-1)\nu_{+}^{0}\left(\mathcal{E}\right)\log_{2}\nu_{+}^{0}\left(\mathcal{E}\right) +n\sum_{\ell=1}^{M-1}\nu^{\ell}\left(\mathcal{E}\right)\log_{2}\nu^{\ell}\left(\mathcal{E}\right)\Bigg],
\end{eqnarray}
where we recall that ${\mathcal{E}}=nE$. 

\begin{figure}
\includegraphics[scale=.24]{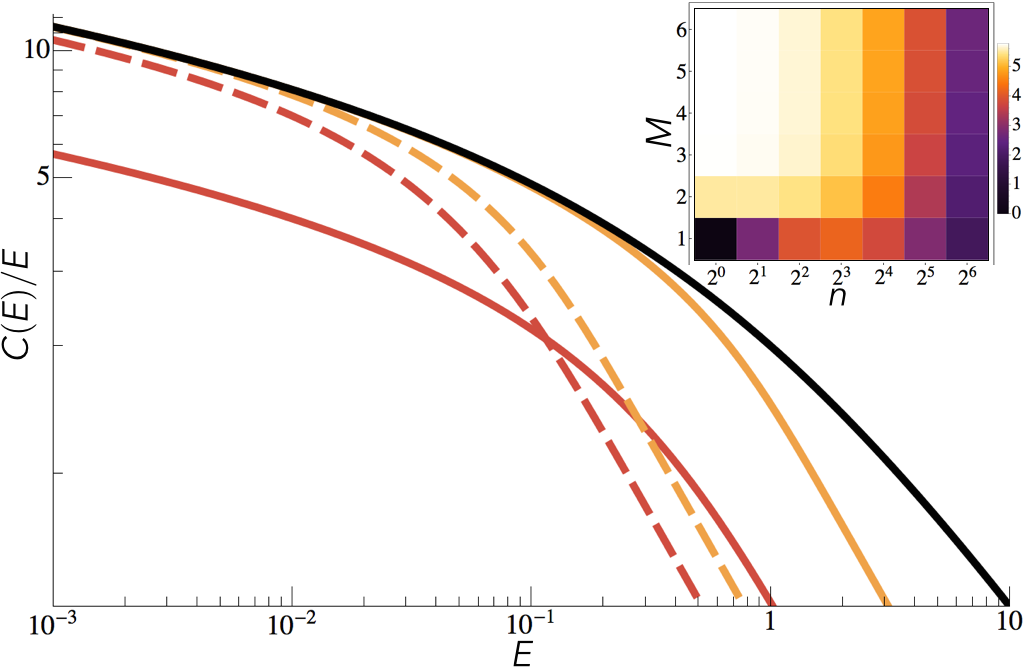}
\caption{Plot (log-log scale) of the optimal rate $R^{(n,M)}_{opt}(E)/E$ per average mode-energy of the PSK Hadamard code of order $M$,  for codewords of length $n=2, 2^{4}$ (respectively solid and dashed lines) and a number of phases $M=1, 4$ (respectively brown/dark-grey and orange/light-grey lines) and the classical capacity per average mode-energy, $C(E)/E$ (black solid line), as a function of the average mode-energy $E$. The quantity $R^{(n,M)}_{opt}(E)/E$ quickly increases at low energy values and for $M>1$ it even saturates the value of $C(E)/E$. For $n>1$ these features appear at lower energy. The inset shows $R^{(n,M)}_{opt}(E)/E$ at $E=0.05$, for several values of the couple $(n,M)$, each represented by a coloured tile. Darker colours are associated to lower values. The general behaviour on this plane seems to favour higher $M$ values at fixed $n$ and lower $n$ values at fixed $M$, except for $M=1$, where the capacity has its peak at $n^{*}(E)=2^{3}$. Also note that for $M>1$ and $n\lesssim n^{*}(E)$ the capacity is approximately constant and equal to its optimal value at that energy, i.e. the light-coloured region of the $(n,M)$ plane. }\label{fig1}
\end{figure}

The quantity $R^{(n,M)}_{opt}(E)$ represents the maximum rate of bits per mode  which one could convey on the channel by means of the code 
 ${\cal H}^{(n)}_M(\alpha)$ when the receiver of the messages is capable of performing optimal measurements~\cite{HolevoBOOK,seq1,seq2,schumawest,polarWildeGuha,NOSTRO}. 
In Fig.~\ref{fig1} we plot $R^{(n,M)}_{opt}(E)/E$ (i.e. the value of  the optimal rate  of ${\cal H}^{(n)}_M(\alpha)$ per average mode-energy) for different values of $M$ and $n$. As a comparison we also report the value of the  classical capacity~\cite{HolevoBOOK,holevo1,holevo2,holevo3} of the channel per average mode-energy $C(E)/E$,  with~\cite{EXACT,YUEN}
\begin{eqnarray} 
C(E)=(E+1)\log_{2}(E+1)-E\log_{2}E \;, \label{CCC}
\end{eqnarray} 
which represents the ultimate rate of the channel attainable when optimizing also the coding structure, as discussed in Sec.~\ref{sec2}
We observe that  for $M>1$, the quantity $R^{(n,M)}_{opt}(E)/E$ considerably increases in the low-energy regime  where it asymptotically saturates the  upper bound $C(E)/E$; unfortunately this increase takes place at lower energy values for larger codewords' length $n$ (see Appendix~\ref{SHOW} for details). In the inset we show the values of $R^{(n,M)}_{opt}(E)/E$, at fixed $E=0.05$, for several values of the couple $(n,M)$, each represented by a coloured tile (darker colours represent lower values). In general, to obtain higher capacities it is clearly convenient to increase $M$ at fixed $n$ and to decrease $n$ at fixed $M$ (the case $M=1$ representing an exception). In particular  for $M>1$,  $R^{(n,M)}_{opt}(E)/E$  seems to remain close to an optimal value for all $n$ below a critical threshold. We conclude that optimal values of the rate $R^{(n,M)}_{opt}(E)$ at fixed energy are obtained for large $M$ and small $n$ values, which motivates us to devise a detection scheme for this family of codes.

\section{Rate of the  PSK Hadamard receiver}\label{secHadRate}
\begin{figure}[t]
\includegraphics[scale=.25]{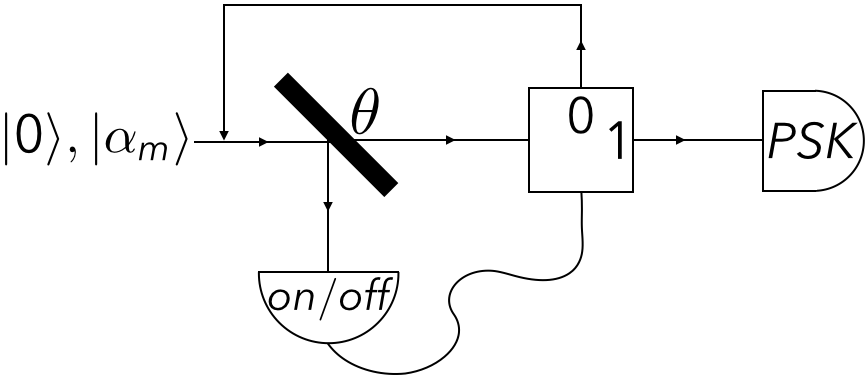}
\caption{Schematic depiction of the single-mode VP detection scheme, which determines whether or not a pulse was present on the given mode and in case of a positive answer determines its phase: the input state, $\ket{\alpha_{m}}$, $m=0,\cdots,M-1$ or the vacuum, is sent through a beam splitter of reflectivity $\eta_{1}=1/N$ (transmissivity $\theta_{1}=1-\eta_{1}$) and its reflected part is measured with an ideal on-off photodetector. If the detector clicks, i.e., ``$1$'', then a pulse must be present and the transmitted part of the state is sent through PSK detection, which identifies the pulse among the $M$ possible ones. If the detector does not click instead, i.e., ``0'', then the received state could be the vacuum, so its transmitted part is sent back to the beam splitter, repeating the same detection procedure with a rescaled reflectivity, $\eta_{2}=\eta_{1}/ \theta_{1}$, so that the fraction of energy incident on the photodetector is always $1/N$. The scheme is iterated $N$ times, with reflectivity $\eta_{p}=\eta_{p-1}/\theta_{p-1}$ at the $p$-th photodetection step, until either a click triggers PSK detection at some iteration step, identifying the phase of the pulse, or the entire energy is exhausted, in which case the receiver guesses that vacuum was sent. The conditional probabilities of detection in the limit $N\rightarrow\infty$, $1-\theta=1/N$, are given by \eqref{DDDlm}. An upper and lower bound on the PSK probability of detection are given in the text (see Eq.~\eqref{multiHel} and Appendix~\ref{naiveProb}).}\label{fig2}
\end{figure}

In this section we want to compute the rate of information transmitted per mode when reading out the PSK Hadamard code with the  PSK Hadamard receiver. 
As we mentioned in Sec.~\ref{sec2} the ultimate advantage of this code is the unitary equivalence between low-energy pulses distributed on several modes by the sender and high-energy ones concentrated on a single mode by the receiver, i.e. respectively the $\ket{v_{k}(\alpha_m)}$ and $\ket{w_{k}(\alpha_m)}$ states of Eq.~(\ref{TRANS2}): the Hadamard transform is able to concentrate the scattered input energy, for a limited number of codewords. The receiver takes advantage of this concentration with the readout operation, which is a separable technique repeated on each mode, labeled Vacuum or Pulse (VP) detection (see Fig.~\ref{fig2}). Its purpose is first to determine whether a pulse is present on that mode then, in case of a positive answer, to determine which among the possible states $\{ \ket{\sqrt{n}\alpha_{m}}; m=0, 1, \cdots, M-1\}$ is the one that was selected by the sender. The simplest case $M=2$ was proposed and employed in Ref.~\cite{Guha2}, while here we discuss its generalization to $M>2$.

The VP scheme can be implemented as shown in Fig.~\ref{fig2}: first we split the unknown received state in two lower-energy copies, by means of a beam splitter of reflectivity $\eta_{1}=1/N$ (transmissivity $\theta_{1}=1-\eta_{1}$), then we measure the reflected copy with an on-off photodetector. Assuming no dark counts and perfect efficiency, the detector can click (``1'' in the figure) only if a pulse was present. In this case, the transmitted copy of the state is employed for PSK detection, which identifies the phase of the pulse among the $M$ possible ones. On the other hand if the detector does not click (``0'' in the figure) we can not exclude that the received state is the vacuum. Hence we repeat the initial procedure on the transmitted copy, sending it back to the beam splitter, measuring its reflected part on the photodetector and applying the same decision rule. Note that we employ a beam splitter with rescaled reflectivity  $\eta_{2}=\eta_{1}/\theta_{1}$, so that the reflected part of the signal at this second step still carries a fraction $1/N$ of the input energy. We iterate this procedure of splitting the state and measuring its reflected part $N$ times, until either a click is registered at some iteration step, triggering PSK detection on the remaining part of the state and the identification of a pulse, or the photodetector never clicks, in which case we guess that the received state was the vacuum.  By properly rescaling the reflectivity at each step  $p$ as $\eta_{p}=\eta_{p-1}/\theta_{p-1}$, we can assure to always extract and measure a fraction $1/N$ of the input energy, so that after $N$ extractions, i.e., no clicks, the energy is exhausted. 
\\
Accordingly the conditional probability of detecting a state $\ket{\alpha_{\ell}}$ if $\ket{\alpha_{m}}$ was sent, $P_{vp}^{(M)}(\ell|m;\mathcal{E},N,\theta)$, is given by the sum, over all values of the step index $p$, of the probability of detecting the first photon at that step and then switching to PSK detection of the $M$ pulses with remaining energy $\theta^{p}\mathcal{E}$. We have:
\begin{align}\label{correctCondN}
P_{vp}^{(M)}(\ell|m;\mathcal{E},N)&=\sum_{p=1}^{N}e^{-\frac{\mathcal{E}}{N}p}\left(1-e^{-\frac{\mathcal{E}}{N}}\right)\nonumber\\
&\cdot P_{psk}^{(M)}\left(\ell|m
;\frac{\mathcal{E}}{N}(N-p)\right),
\end{align}
where $e^{-\frac{\mathcal{E}}{N}}$ is the conditional probability of registering no click at any photodetection step if a pulse was present, while $P_{psk}^{(M)}\left(\ell|m;\theta^{p}\mathcal{E}\right)$ represents the PSK-detection probability of guessing $\ket{\alpha_{\ell}}$ if $\ket{\alpha_{m}}$ was sent, after the photodetector clicked at the $p$-th step; its specific form will be discussed after computing the rate of the receiver. If instead the photodetector never clicks, we guess that the vacuum was present, even though it may have actually been a pulse, making an error with probability
\begin{align}
\label{err0CondN}
\mathbb{P}_{vp}^{(M)}(v|m;\mathcal{E},N)=e^{-\mathcal{E}}.
\end{align}
Equation~(\ref{correctCondN}) can be evaluated as an integral in the limit of infinite splitting steps, $N\rightarrow\infty$, as detailed in Appendix~\ref{AppB}. The final result is
\begin{align}
\label{DDDlm}\begin{aligned}
\mathbb{P}_{vp\text{-}psk}^{(M)}\left(\ell|m;\mathcal{E}\right)&=\lim_{N\rightarrow\infty} P_{vp}^{(M)}(\ell|m;\mathcal{E},N)\\
&=\int_{e^{-\mathcal{E}}}^{1}dt~P_{psk}^{(M)}(\ell|m;\mathcal{E}+\ln t).
\end{aligned}
\end{align}\\
Eventually the detection of a whole Hadamard codeword $\ket{w_{k}(\alpha_{m})}$ is carried out by applying VP to each mode. With this method a codeword with pulse on mode $k$ can be misinterpreted only as one of the other $M-1$ codewords living on the same mode, since all other modes are occupied 
by the vacuum, which with unit probability never clicks. The only additional source of error is when the pulse on the $k$-th mode does not click, in which case no codeword can be identified. Accordingly the rate of the whole receiver is computed in terms of the mutual information of the classical input/output random variables induced by the quantum encoding/decoding operations, respectively:  $x\in X=\{(k,m)\}_{k=0,m=0}^{n-1,M-1}$, determined by the index of the mode where the pulse is present, $k$, and the phase index of the pulse, $m$, and $y\in Y=X\cup\{Err\}$, with an additional error outcome associated to the case where VP does not click on any mode. The input probability distribution is uniform, $P_{X}(x)=1/(nM)$, while the conditional output one is:
\begin{equation}
P_{Y|X}(y|x)=\begin{cases}
\mathbb{P}_{vp\text{-}psk}^{(M)}(v|m_{x};\mathcal{E})&\text{ if }y=Err,\\
\mathbb{P}_{vp\text{-}psk}^{(M)}(m_{y}|m_{x};\mathcal{E})&\text{ if }k_{y}=k_{x},\\
0&\text{ otherwise}.
\end{cases}
\end{equation}
The rate of the PSK Hadamard receiver with average received energy per mode $E$ is then given by:
\begin{equation}\begin{aligned}\label{hadRate}
R^{(n,M)}_{had}(E)&=\sum_{m_{x},m_{y}=0}^{M-1}\frac{\mathbb{P}_{vp\text{-}psk}^{(M)}(m_{y}|m_{x};\mathcal{E})}{Mn}\\
&\cdot\log_{2}\left(\frac{M n~\mathbb{P}_{vp\text{-}psk}^{(M)}(m_{y}|m_{x};\mathcal{E})}{\sum_{m_{x'}=0}^{M-1} \mathbb{P}_{vp\text{-}psk}^{(M)}(m_{y}|m_{x};\mathcal{E})}\right).
\end{aligned}\end{equation}
Let's now discuss the specific form of $P_{psk}^{(M)}(\ell|m;\mathcal{E})$, which determines $\mathbb{P}_{vp\text{-}psk}^{(M)}\left(m_{y}|m_{x};\mathcal{E}\right)$ and hence the rate $R^{(n,M)}_{had}(E)$. For the case $M=2$, it was shown that Dolinar detection \cite{Dol,holDol,ImplProj,Multicopy,DolMultiplexed,DolExp}, based on splitting, conditional signal nulling and fast feedforward, attains the optimal success probability (Helstrom bound)~\cite{Hel} of discriminating the binary coherent states $|\pm \alpha\rangle$ (see Ref.~\cite{DolExp} for a recent proof-of-principle experimental demonstration and Ref.~\cite{DolImp} for an evaluation of its performance in the case of imperfect detection). For the general case, i.e., $M>2$, a variety of schemes has been proposed in the past \cite{bondurant,izumi,guhaDol,becerra1,becerra2,marquardt}, generalizing that of Dolinar. The large literature on the subject seems to suggest that this generalized Dolinar detection can get close to the PSK Helstrom bound \cite{multiHel1,multiHel2} on the discrimination of $M$ symmetrically-distributed coherent states of fixed intensity, in particular surpassing the performance obtained by classical detection techniques. Nevertheless a proof of this fact like that of Refs.~\cite{Dol,holDol,ImplProj} for the two-state Dolinar scheme still lacks. In order to have a fair comparison with the $M=2$ case and evaluate the best performance of our receiver, in the following we report both the detection efficiency one could reach by employing a optimal measurement that saturates the PSK Helstrom probability of discrimination \cite{multiHel2} at the second step, 
\begin{align}\begin{aligned}\label{multiHel}
P_{hel}^{(M)}(\ell|m;\mathcal{E})=\abs{\frac{1}{M}\sum_{j=0}^{M-1}e^{-i \frac{2\pi}{M} j(\ell-m)}\sqrt{\lambda_{j}(\mathcal{E})}}^{2},
\end{aligned}\end{align}
with $\lambda_{j}(\mathcal{E})$ defined in \eqref{mpskEigen}, and the detection efficiency one can attain by employing a very naive but realistic generalization of the Dolinar scheme, $P_{real}^{(M)}(\ell|m;\mathcal{E})$, based on the same splitting method of the VP detector and sequentially nulling one of the hypoteses, discarding it if a click is registered (see Appendix~\ref{naiveProb} for its detailed form). Hence we have an upper and lower bound for the PSK probability of discrimination, i.e., $P_{hel}^{(M)}(\ell|m;\mathcal{E})\geq P_{psk}^{(M)}(\ell|m;\mathcal{E})\geq P_{real}^{(M)}(\ell|m;\mathcal{E})$; when substituting these bounds in \eqref{DDDlm}, we obtain two new expressions for the conditional probability of detection, $\mathbb{P}_{vp\text{-}hel}^{(M)}(\ell|m;\mathcal{E})=\int_{e^{-\mathcal{E}}}^{1}dt~P_{hel}^{(M)}(\ell|m;\mathcal{E}+\ln t)$  and $\mathbb{P}_{vp\text{-}real}^{(M)}(\ell|m;\mathcal{E})=\int_{e^{-\mathcal{E}}}^{1}dt~P_{real}^{(M)}(\ell|m;\mathcal{E}+\ln t)$, which, through equation \eqref{hadRate}, provide a correspondent upper and lower bound for the rate of the Hadamard receiver, i.e., $R^{(n,M)}_{hel}(E)\geq R^{(n,M)}_{had}(E)\geq R^{(n,M)}_{real}(E)$, with
\begin{equation}\begin{aligned}\label{helRate}
R^{(n,M)}_{hel}(E)&=\sum_{m_{x},m_{y}=0}^{M-1}\frac{\mathbb{P}_{vp\text{-}hel}^{(M)}(m_{y}|m_{x};\mathcal{E})}{Mn}\\
&\cdot\log_{2}\left(\frac{M n~\mathbb{P}_{vp\text{-}hel}^{(M)}(m_{y}|m_{x};\mathcal{E})}{\sum_{m_{x'}=0}^{M-1} \mathbb{P}_{vp\text{-}hel}^{(M)}(m_{y}|m_{x};\mathcal{E})}\right),
\end{aligned}\end{equation}
and $R^{(n,M)}_{real}(E)$ explicitly given in Appendix~\ref{naiveProb} for $M=3,4$.
 In particular the upper bound is saturated in the case $M=2$, while for $M>2$ there are optimized schemes which get close to it at low energy (see Ref.~\cite{marquardt} for a detailed description). \\
It is important to have a separable communication scheme to compare with, based on the same constellation of single-mode states. It seems reasonable to send one of the $M$ symmetric coherent states on each mode and read them out with generalized Dolinar detection. We call the latter a  PSK Separable receiver and its rate is:
\begin{equation}\begin{aligned}
R^{(M)}_{sep}(E)&=\sum_{m_{x},m_{y}=0}^{M-1}\frac{P_{hel}^{(M)}(m_{y}|m_{x};E)}{M}\\
&\cdot\log_{2}\left(\frac{M~P_{hel}^{(M)}(m_{y}|m_{x};E)}{\sum_{m_{x'}=0}^{M-1} P_{hel}^{(M)}(m_{y}|m_{x};E)}\right).
\end{aligned}\end{equation} 
\begin{figure}[t]
\includegraphics[scale=0.24]{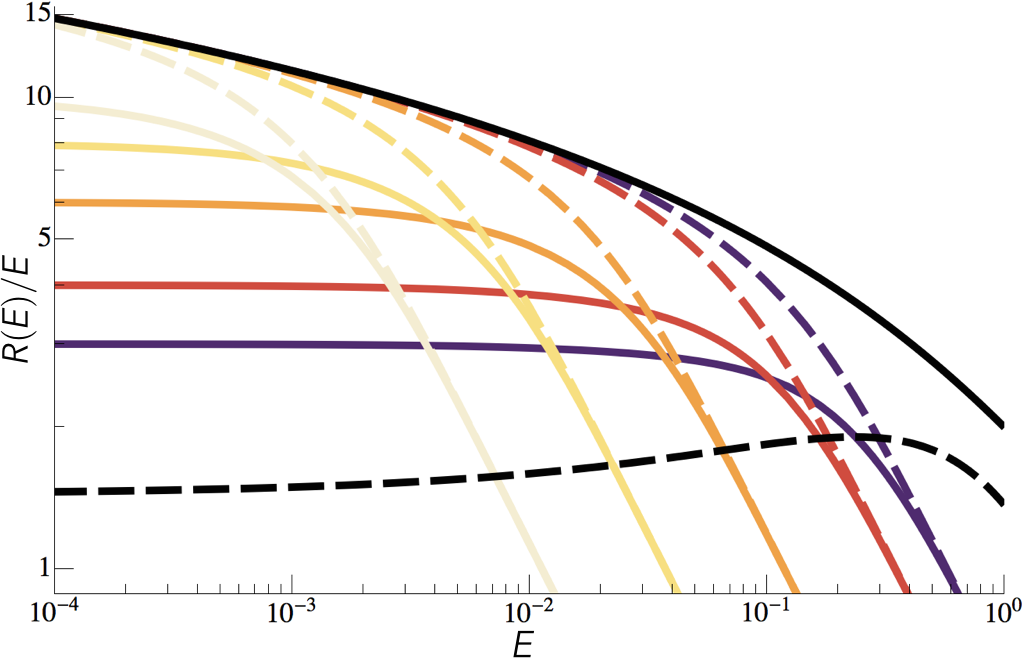}
\caption{Plot (log-log scale) of the upper-bound of the Hadamard rate per average mode-energy, $R_{hel}^{(n,M)}(E)/E$, for $M=3$ phases and codewords of length $n=2^{i}$, $i=3,4,6,8,10$ (solid lines with colours from purple/dark grey to light brown/light grey), the corresponding optimal rate per average mode-energy, $R_{opt}^{(n, M)}(E)/E$, for the same values of $n,M$ (dashed lines with colours from purple/dark grey to light brown/light grey), the Separable rate per average mode-energy, $R_{Sep}^{(M)}(E)/E$, for the same number of phases $M=3$ (black dashed line) and the classical capacity of the channel, $C(E)/E$ (black solid line), as a function of the average mode-energy $E$. Observe that the Hadamard rates rise over their separable counterpart at low energy, only to level off at a plateau later on; this effect is shifted towards lower energies and higher plateau values for larger codewords' length $n$. Note also that, for any $n$, the Hadamard rate attains its optimal value $R_{opt}^{(n,M)}(E)$ at high energy, but detaches towards the plateau when the latter attains the classical capacity of the channel.}\label{fig3}
\end{figure}

In Fig.~\ref{fig3} we show the Hadamard rate, with Helstrom PSK detection, per average mode-energy as a function of the average mode-energy, $R^{(n,M)}_{hel}(E)/E$ vs. E, for $M=3$ phases and several values of the codewords' length $n$, along with $R^{(M)}_{sep}(E)/E$ and $R^{(n,M)}_{opt}(E)/E$, also for $M=3$ and the same $n$ values, and $C(E)/E$. We observe that all Hadamard rates, for any value of $n$, surpass the Separable rate below a certain low-energy threshold, $E^{*}(n)<1$, then leveling off at a plateau for smaller energy. As $n$ increases, this energy threshold $E^{*}(n)$ lowers, while the plateau value grows. Moreover the Hadamard rate attains its optimal value at high energy $E>E^{*}(n)$, but detaches from it and from the classical capacity of the channel when leveling off at the low-energy plateau. Still note that, if we take the envelope of $R^{(n,M)}_{hel}(E)$ for fixed $M$ and all allowed values of $n$ in a given interval $\mathcal{N}$, i.e., \begin{eqnarray}\label{helEnv}
R^{(\mathcal{N},M)}_{hel}(E)=\max_{n\in \mathcal{N}}R^{(n,M)}_{hel}(E)\;, \end{eqnarray} we are able to get closer to $C(E)$ as the energy decreases.
\begin{figure}[t]
\includegraphics[scale=0.24]{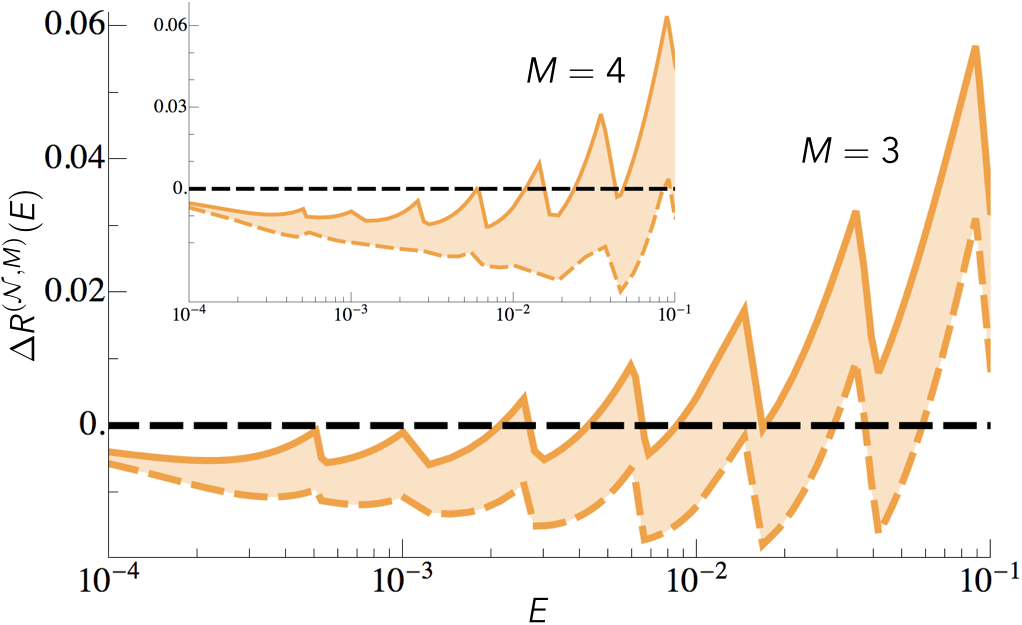}
\caption{Plot (log-linear scale) of $\Delta R_{hel}^{(\mathcal{N}, M)}(E)$ (orange/light-grey solid line) and $\Delta R_{real}^{(\mathcal{N},M)}(E)$ (orange/light-grey dashed line) as a function of the average energy per mode $E$ for $M=3$ and, in the inset, $M=4$. The shaded region between the two curves represents all rate values achievable by Hadamard coding and VP detection. The black dashed line of constant value $0$ represents the reference quantity $\mathcal{R}_{hel}^{(\mathcal{N}, 2)}(E)\equiv\mathcal{R}_{had}^{(\mathcal{N}, 2)}(E)$. Observe that the Hadamard rate with $M=3, 4$ phases beat both the that with $M=2$ phases in an interval of energy $E\sim[4\cdot10^{-3},10^{-1}]$, with a gain of up to $6\%$ over the rate at $M=2$. Hence the design of better PSK detection techniques, achieving the Helstrom bound for PSK states would provide access to the best communication rates so far in the low-energy regime.}\label{fig4}
\end{figure}\\
\begin{figure}[t]
\includegraphics[scale=0.24]{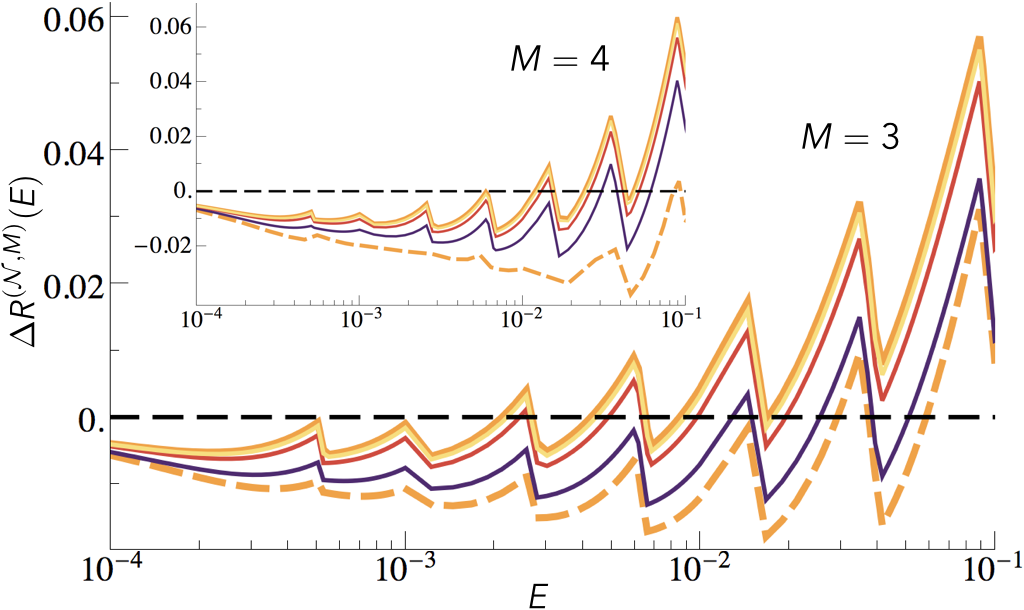}
\caption{Plot (log-linear scale) of the same rates of Fig.~\ref{fig4} and of those achievable by the Helstrom-Hadamard receiver with a finite number of splitting steps $N=10, 30, 100$ (respectively from purple/dark-grey to yellow/light-grey solid lines), based on \eqref{correctCondN}. For both numbers of phases $M=3,4$, already $30$ splitting steps (red/middle lines) achieve a rate close to the one obtained in the infinite-$N$ limit (orange/light-grey solid curve).}\label{fig5}
\end{figure}\\

Eventually we want to compare the performance of Hadamard rates for $M>2$  with that for the case $M=2$, discussed in Ref.~\cite{Guha2}. In Fig.~\ref{fig4} we show the difference between the upper-bound Hadamard envelopes \eqref{helEnv} for $M=3,4$ and that for $M=2$, relative to the latter, i.e., $\Delta R^{(\mathcal{N},M)}_{hel}(E)=\left[R^{(\mathcal{N},M)}_{hel}(E)-R^{(\mathcal{N},2)}_{hel}(E)\right]/R^{(\mathcal{N},2)}_{hel}(E)$, with $\mathcal{N}=\{2^{i},i=1,\cdots,10\}$. The same quantity is plotted also for the lower bound of the Hadamard rate, discussed in Appendix~\ref{naiveProb}, i.e., $\Delta R^{(\mathcal{N},M)}_{real}(E)$, so that the shaded region between the two curves represents all rates achievable by the Hadamard code with VP detection. We observe an advantage of the schemes with a higher number of phases in the energy region $E\in[4\cdot10^{-3},10^{-1}]$. For higher energy values the Separable technique becomes superior, since the classical capacity allows for more codewords to be used, while for lower energy the Hadamard rate for $M=2$ performs better, since distinguishability of the signals becomes crucial. We conclude that the practical Hadamard receiver with more than two phases has relevant rate, with a gain of up to $6\%$ with respect to previous proposals, in a crossover intensity region where separable techniques start performing worse than joint ones. In Fig.~\ref{fig5} we show the same plots with the  addition of three curves, corresponding to the maximum rate achievable by the Hadamard receiver with a finite number of splitting steps $N=10, 30, 100$, i.e., employing the conditional probability \eqref{correctCondN} instead of \eqref{DDDlm}. It can be seen that a reasonable approximation of the infinite-$N$ limit is obtained already for $N=30$, independently of the number of phases $M=3,4$.

\section{Conclusions}\label{secConc}
We have shown that the Hadamard receiver can reach higher rates than previously thought, by adding up to four phases to the coherent pulses and performing generalized Dolinar detection. Moreover the optimal rate of the Hadamard code attains the classical-quantum capacity of the channel faster when increasing the number of phases. This suggests that Hadamard coding is a good block-coding technique at low energy, indeed the best known so far, but its rate is still affected by a huge detection gap. Nevertheless the receiver seems structured to maximize the recovery of information, gathering it on one mode, then reading it out with Helstrom-optimal success probability. We conclude that better low-energy, joint-detection receivers may be designed in the future by uniting the crucial ingredient of the Hadamard one, i.e., sending few information per mode, few codewords per number of modes, as allowed by the classical-quantum capacity at those energies, with more refined multi-mode detection techniques.  

\appendix

\section{Determining $\hat{U}^{(n)}_{had}$} \label{APPA}
It is easy to show that the operator $\hat{U}^{(n)}_{had}$ has symplectic representation $H_{n}/\sqrt{n}$. Indeed consider the  gaussian $n$-mode complex displacement operator, $\hat{D}\left(\underline{\alpha}\right)=\exp\left[\underline{\hat{a}}^{\dagger}\cdot\underline{\alpha}- \text{h.c.}\right]$, where $\underline{\hat{a}}=(\hat{a}_{0},\cdots,\hat{a}_{n-1})^{T}$ is the column vector of bosonic annihilation operators for each mode, $\underline{a}^{\dagger}$ its conjugate transpose and $\underline{\alpha}=(\alpha_{(0)},\cdots,\alpha_{(n-1)})^{T}$ the corresponding vector of complex mean values. We can then express any received state as $\ket{v_{k}(\alpha)}=\hat{D}\left(\alpha~\underline{h}_{k}\right)\ket{0}^{\otimes n}$, with $\underline{h}_{k}=\left(\left(H_{n}\right)_{1,k},\cdots,\left(H_{n}\right)_{n-1,k}\right)^{T}$ and when applying the passive unitary we have:
\begin{eqnarray}\label{Transform}
&\hat{U}^{(n)}_{had}&\ket{v_{k}(\alpha)}=\hat{D}\left(\frac{\alpha}{\sqrt{n}} H_{n}\underline{h}_{k}\right)\ket{0}^{\otimes n}\nonumber\\
&&=\hat{D}\left(\sqrt{n}\alpha(0,\cdots,0,1_{(k)},0,\cdots,0)\right)\ket{0}^{\otimes n}\\
&&=\ket{w_{k}(\alpha)}\nonumber,\end{eqnarray}
where the second equality follows from the orthogonality property of $H_{n}$.

\section{Spectral decomposition of $\Gamma^{(e)}$}\label{AppA}
Here we show that the Gram matrix $\Gamma^{(e)}$ of the codewords $\ket{e_{k}^{0}}$, $k=0,\cdots,n-1$, is diagonalized by a unitary $V=H_{n}/\sqrt{n}$, proportional to the Hadamard matrix of order $n$ (see Sec.~\ref{sec2}) and compute its eigenvalues. Consider the matrix product
\begin{eqnarray}\label{gram}
&\frac{1}{n}(H_{n}\Gamma^{(e)}H_{n}&)_{\ell,k}=\frac{1}{n}\sum_{i,j=0}^{n-1}\left(H_{n}\right)_{\ell,i}(\Gamma^{(e)})_{i,j}\left(H_{n}\right)_{j,k}\nonumber\\
&&=\delta_{\ell k}+\frac{M e^{-\mathcal{E}}}{n\lambda_{0}(\mathcal{E})}\sum_{i,j\neq i}\left(-1\right)^{\ell\cdot i+k\cdot j},
\end{eqnarray}
where we have applied the definition \eqref{HadTerm} of $H_{n}$ and employed the fact that the off-diagonal terms of $\Gamma^{(e)}$ are all equal, as implied by \eqref{eOverlaps}. The last term in the latter equation can be simplified further, by computing the sum
\begin{align}
\begin{aligned}
\sum_{j=0}^{n-1}(-1)^{k\cdot j}&=\prod_{t=1}^{\log_{2}n}\sum_{j_{t}=0,1}(-1)^{k_{t}j_{t}}\\
&=\prod_{t=1}^{\log_{2}n}\left(2\delta_{k_{t},0}\right)=n\delta_{k,0}.
\end{aligned}
\end{align}
We have then:
\begin{align}\begin{aligned}
\sum_{i,j\neq i}\left(-1\right)^{\ell\cdot i+k\cdot j}&=\sum_{i=0}^{n-1}(-1)^{\ell\cdot i}\left(n\delta_{k,0}-(-1)^{k\cdot i}\right)\\
&=n^{2}\delta_{\ell,0}\delta_{k,0}-\prod_{t=1}^{\log_{2}n}\left(2\delta_{\ell_{t}+k_{t},0}\right)\\
&=n^{2}\delta_{\ell,0}\delta_{k,0}-n\delta_{\ell,k},
\end{aligned}\end{align}
where in the last equality we have used the fact that the binary sum $\ell_{t}+k_{t}$ can be zero also if both bits are equal to one, so that $\delta_{\ell_{t}+k_{t},0}=\delta_{\ell_{t},k_{t}}$. Inserting this expression in Eq. \eqref{gram} we obtain:
\begin{equation}
\frac{1}{n}(H_{n}\Gamma^{(e)}H_{n})_{\ell,k}=\delta_{\ell,k}\left(1+(n\delta_{\ell,0}-1)\frac{M e^{-\mathcal{E}}}{\lambda_{0}(\mathcal{E})}\right),
\end{equation}
confirming that $V=H_{n}/\sqrt{n}$ diagonalizes $\Gamma^{(e)}$, with eigenvalues $\mu_{\ell}(\mathcal{E})=1+(n\delta_{\ell,0}-1)\frac{M e^{-\mathcal{E}}}{\lambda_{0}(\mathcal{E})}$.

\section{Analytical approximation of the optimal Hadamard rate for low energy}\label{SHOW} 
Here we  show that  the optimal Hadamard rate $R_{opt}^{(n,M)}(E)$ of Eq.~(\ref{mpskHolCap})  attains the classical capacity (\ref{CCC}) of the channel for low energy, for any codewords' length $n$ and for any  $M>1$. In order to show this  we take the low-energy expansion $\mathcal{E}\ll 1$ up to first order. First we notice that  the coefficients \eqref{mpskEigen} give
\begin{equation}
\begin{aligned}
\lambda_{\ell}\left(\mathcal{E}\right)&\simeq \sum_{h=0}^{M-1}\left(1-\left(1-e^{i\frac{2\pi}{M}h}\right)\mathcal{E}\right)e^{-i\frac{2\pi}{M}\ell h}\\
&=\left(1-\mathcal{E}\right)M\delta_{\ell,0}+\mathcal{E}M\delta_{\ell,1}.
\end{aligned}
\end{equation}
Hence the eigenvalues of $\bar{\rho}^{(+)}_{PPM}$, Eq. \eqref{rho+}, and $\bar{\rho}_{PPM}^{(0)}$, Eq. \eqref{rho0Fin}, yield respectively:
\begin{align}
&\begin{aligned}
\nu^{\ell>0}(\mathcal{E})\simeq \delta_{\ell,1}\mathcal{E}/n\;, \end{aligned}\\
&\begin{aligned}
\nu_{0}^{0}\left(\mathcal{E}\right)\simeq\frac{(1-\mathcal{E})M+(n-1)M(1-\mathcal{E})}{Mn}=1-\mathcal{E},
\end{aligned}\\
&\begin{aligned}
\nu_{+}^{0}(\mathcal{E})\simeq\frac{(1-\mathcal{E})M-M(1-\mathcal{E})}{Mn}=0.
\end{aligned}
\end{align}
Inserting the previous expressions in \eqref{mpskHolCap} we then obtain: 
\begin{equation}\begin{aligned}\label{hadCapApp}
R_{opt}^{(n,M)}(E)&\simeq-\frac{(1-\mathcal{E})\log_{2}(1-\mathcal{E})+\mathcal{E}\log_{2}(\frac{\mathcal{E}}{n})}{n}\\
&\simeq E-E\log_{2}E.
\end{aligned}\end{equation}
which coincides with the expansion of the classical capacity~(\ref{CCC}), meaning that the two quantities are the same at sufficiently low energy $\mathcal{E}\ll 1$, i.e., $E\ll1/n$. Also note that for $M=1$ there is no contribution from the eigenvalues $\nu^{\ell>0}$, so that the second term in \eqref{hadCapApp} is missing. This explains why  for this value, $R_{opt}^{(n,M)}(E)$ does not saturate the bound $C(E)$ even at high values of $n$.

\section{Computation of the VP conditional detection probability}\label{AppB}
Here we show that the VP conditional probability of detection, Eq.~(\ref{correctCondN}), can be written as an integral in the limit $N\rightarrow\infty$. It is sufficient to define the variable $x_{p}=p\mathcal{E}/N\in[\mathcal{E}/N,\mathcal{E}]$, whose increment is $\Delta x=\mathcal{E}/N$, infinitesimal in the large-$N$ limit. Hence the conditional probability (\ref{correctCondN}) can be written at the first order in $\Delta x$ as
\begin{eqnarray}
&P_{vp}^{(M)}(\ell|m;\mathcal{E}&,N)\simeq\sum_{p=1}^{N} e^{-x_{p}} \Delta x~P_{psk}^{(M)}\left(\ell|m;\mathcal{E}-x_{p}\right)\nonumber\\
&&\underset{N\rightarrow\infty}{\longrightarrow}\int_{0}^{\mathcal{E}}dx e^{-x} P_{psk}^{(M)}\left(\ell|m;\mathcal{E}-x\right),
\end{eqnarray}
which gives Eq. \eqref{DDDlm} after a change of variable $t=e^{-x}$.

\section{Computation of the realistic PSK detection probability and of the corresponding Hadamard rate}\label{naiveProb}
As discussed in Sec.~\ref{secHadRate}, it is not known if a detection technique like that of Dolinar achieves the Helstrom-optimal probability of discrimination between $M>2$ coherent states. Hence here we compute the detection probability of PSK signals employing a naive yet realistic detection scheme. Better detection schemes can be developed by refining this one, as discussed in Ref.~\cite{marquardt}.
We employ a device similar to that of Fig. \ref{fig2}, where the reflected part of the state after the beam-splitter is subject to a displacement of $-\alpha_{0}$ before photodetection. In this way we perfectly null the state $\ket{\alpha_{0}}$, if it was present, and can exclude its presence, if a click is registered. When this happens we proceed with the PSK detection of $M-1$ signals. For $M=3$ in particular, this second step will be ordinary Dolinar detection, which is Helstrom-optimal, while for a higher number of pulse phases $M$, this procedure will establish a hierarchy of subsequent realistic PSK detections, excluding one state at each stage, all the way down to $M=2$ remaining states.  Accordingly for $M=3$ and in the limit of infinite splitting steps $N\rightarrow\infty$, the conditional probability of guessing state $\ket{\alpha_{\ell}}$ if $\ket{\alpha_{m}}$ was sent, $P_{real}^{(M)}(\ell|m;\mathcal{E})$, is:
\begin{align}
&P_{real}^{(3)}(0|0;\mathcal{E})=1,\quad P_{real}^{(3)}(1|0;\mathcal{E})=0,\\
&P_{real}^{(3)}(0|1;\mathcal{E})=\abs{\dbraket{0}{\alpha_{1}-\alpha_{0}}}^{2}=e^{-3\mathcal{E}},\\
&\begin{aligned}P_{real}^{(3)}(1|1;\mathcal{E})&=\int_{0}^{\abs{\alpha_{1}-\alpha_{0}}^{2}}dx~e^{-x}\\
&\cdot P_{hel}^{(2)}\left(1|1;\frac{\abs{\alpha_{1}-\alpha_{2}}^{2}-x}{2}\right)\\
&=\int_{e^{-3\mathcal{E}}}^{1}dy~P_{hel}^{(2)}\left(1|1;\frac{3\mathcal{E}+\ln y}{2}\right),\end{aligned}\\
&\begin{aligned}P_{real}^{(3)}(2|1;\mathcal{E})&=\int_{e^{-3\mathcal{E}}}^{1}dy~P_{hel}^{(2)}\left(2|1;\frac{3\mathcal{E}+\ln y}{2}\right)\\
&=1-P_{real}^{(3)}(0|1;\mathcal{E})-P_{real}^{(3)}(1|1;\mathcal{E}),\end{aligned}
\end{align}
where the integrals have been obtained by means of the same method of Appendix~\ref{AppB}, with the rescaling $x\rightarrow x\cdot\abs{\alpha_{1}-\alpha_{0}}^{2}/\mathcal{E}$, which takes into account the energy of the states after the nulling displacement. Given the symmetry of the states, the remaining conditional probabilities can be obtained by switching $1\leftrightarrow 2$ in the previous equations.
After substituting the previous expressions in Eq. \eqref{DDDlm} we obtain the VP conditional probability of detection when this realistic PSK detection scheme is employed. Eventually the Hadamard rate for $M=3$ can be computed along the same lines of  Sec.~\ref{secHadRate} and is equal to
\begin{eqnarray}
&&R^{(n,3)}_{real}(E)=h\left[\frac{1-e^{-3\mathcal{E}}}{3n}\right]+2h\left[\frac{2-3e^{-\mathcal{E}}+e^{-3\mathcal{E}}}{6n}\right]\nonumber\\
&&-\frac{1}{3} h\left[1-e^{-\mathcal{E}}\right]-\frac{2}{3}\Bigg\{h\left[\frac{e^{-\mathcal{E}}-e^{-3\mathcal{E}}}{2}\right]+h\left[\mathbb{P}_{vp\text{-}real}^{(3)}(1|1;\mathcal{E})\right]\nonumber\\
&&+h\left[\frac{2-3e^{-\mathcal{E}}+e^{-3\mathcal{E}}}{2}-\mathbb{P}_{vp\text{-}real}^{(3)}(1|1;\mathcal{E})\right]\Bigg\},\end{eqnarray}
where $h[P]=-P\log_{2}P$, $P\in[0,1]$, is the entropy of a single probability value and, following Eq. \eqref{DDDlm}, we have defined the conditional probability of correctly identifying the state $\ket{\alpha_{1}}$ with VP and realistic PSK detection for $M=3$ as $\mathbb{P}_{vp\text{-}real}^{(3)}(1|1;\mathcal{E})=\int_{e^{-\mathcal{E}}}^{1}dt~P_{real}^{(3)}(1|1;\mathcal{E}+\ln t)$.

Similarly, for $M=4$ we have a two-stage hierarchy of realistic PSK detection where we first null the state $\ket{\alpha_{0}}$. If a click is registered, we are left with three states to discriminate and can resort to the case previously studied, though with a different symmetry between the states. In this case, we null the state $\ket{\alpha_{2}}$, which is equidistant from the other two remaining ones. Hence the conditional probability of realistic PSK detection reads out:
\allowdisplaybreaks\begin{eqnarray}
&P_{real}^{(4)}(0|0;\mathcal{E})&=1,\quad P_{real}^{(4)}(1,2,3|0;\mathcal{E})=0,\\
&P_{real}^{(4)}(0|2;\mathcal{E})&=\abs{\dbraket{0}{\alpha_{2}-\alpha_{0}}}^{2}=e^{-4\mathcal{E}},\\
&P_{real}^{(4)}(0|1;\mathcal{E})&=e^{-2\mathcal{E}},\quad P_{real}^{(4)}(1|2;\mathcal{E})=0,\\
&P_{real}^{(4)}(2|2;\mathcal{E})&=\int_{0}^{\abs{\alpha_{2}-\alpha_{0}}^{2}}dx~e^{-x}\cdot1=1-e^{-4\mathcal{E}},\\
&P_{real}^{(4)}(2|1;\mathcal{E})&=\int_{0}^{\abs{\alpha_{1}-\alpha_{0}}^{2}}dx~e^{-x} e^{-(\abs{\alpha_{1}-\alpha_{2}}^{2}-x)}\nonumber\\
&&=\int_{e^{-2\mathcal{E}}}^{1}dt~e^{-(2\mathcal{E}+\ln t)}=e^{-2\mathcal{E}}2\mathcal{E},\\
&P_{real}^{(4)}(1|1;\mathcal{E})&=\int_{0}^{\abs{\alpha_{1}-\alpha_{0}}^{2}}dx~e^{-x}\nonumber\\
&&\cdot\int_{0}^{(|\alpha_{1}-\alpha_{2}|^{2}-x)}dx'~e^{-x'}\\
&&\cdot P_{hel}^{(2)}\left(1|1;\frac{(\abs{\alpha_{1}-\alpha_{3}}^{2}-x-x')}{2}\right)\nonumber\\
&={\displaystyle \int_{e^{-2\mathcal{E}}}^{1}}dt&\int_{e^{-(2\mathcal{E}+\ln t)}}^{1}dt'P_{hel}^{(2)}\left(1|1;2\mathcal{E}+\ln(t t')\right),\nonumber\\
&P_{real}^{(4)}(3|1;\mathcal{E})&=1-P_{real}^{(4)}(0|1;\mathcal{E})-P_{real}^{(4)}(2|1;\mathcal{E})\nonumber\\
&&-P_{real}^{(4)}(1|1;\mathcal{E}).
\end{eqnarray}
As before, for symmetry reasons, the remaining conditional probabilities can be obtained by switching the indexes $1\leftrightarrow3$.
Eventually the Hadamard rate for $M=4$ is given by
\begin{eqnarray}
&R_{real}^{(4)}(E)&=h\left[\frac{3+4e^{-\mathcal{E}}-6e^{-2\mathcal{E}}-e^{-4\mathcal{E}}}{12n}\right]+\nonumber\\
\nonumber\\
&&+h\left[\frac{3+8e^{-\mathcal{E}}-12e^{-2\mathcal{E}}+e^{-4\mathcal{E}}}{12n}\right]\nonumber\\
&&+2h\left[\frac{1-4e^{-\mathcal{E}}+(3+2\mathcal{E})e^{-2\mathcal{E}}}{4n}\right]\nonumber\\
&&-\frac{1}{4n}\Bigg\{h\left[1-e^{-2\mathcal{E}}\right]+h\left[\frac{e^{-\mathcal{E}}-e^{-4\mathcal{E}}}{3}\right]\\
&&+h\left[\frac{3-4e^{-\mathcal{E}}+e^{-4\mathcal{E}}}{3}\right]\Bigg\}\nonumber-\frac{1}{2n}\Bigg\{h\left[e^{-\mathcal{E}}-e^{-2\mathcal{E}}\right]\\
&&+h\left[2(e^{-\mathcal{E}}-(1+\mathcal{E})e^{-2\mathcal{E}})\right]+h\left[\mathbb{P}_{vp-real}^{(4)}(1|1;\mathcal{E})\right]\nonumber\\
&&+h\left[1-4e^{-\mathcal{E}}+(3+2\mathcal{E})e^{-2\mathcal{E}}-\mathbb{P}_{vp-real}^{(4)}(1|1;\mathcal{E})\right]\Bigg\},\nonumber
\end{eqnarray}
where we have defined the conditional probability of correctly identifying the state $\ket{\alpha_{1}}$ with VP and realistic PSK detection for $M=4$ as $\mathbb{P}_{vp\text{-}real}^{(4)}(1|1;\mathcal{E})=\int_{e^{-\mathcal{E}}}^{1}dz~P_{real}^{(4)}(1|1;\mathcal{E}+\ln z)$.

\end{document}